\documentclass[9pt,conference]{IEEEtran}
\IEEEoverridecommandlockouts

\usepackage{cite}
\usepackage{amsmath,amssymb,amsfonts}
\usepackage{algorithmic}
\usepackage{graphicx}
\usepackage{textcomp}
\usepackage{xcolor}

\usepackage{comment}
\usepackage{booktabs}
\usepackage{multirow}
\usepackage{listings}
\usepackage{subcaption}
\usepackage{amssymb}
\usepackage{bm}

\def\BibTeX{{\rm B\kern-.05em{\sc i\kern-.025em b}\kern-.08em
    T\kern-.1667em\lower.7ex\hbox{E}\kern-.125emX}}
\begin{document}

\title{
Pre-training with Synthetic Patterns for Audio
}

\author{\IEEEauthorblockN{1\textsuperscript{st} Yuchi Ishikawa}
\IEEEauthorblockA{\textit{LY Corporation / Keio University} \\
Tokyo, Japan \\
yuchi.ishikawa@lycorp.co.jp}
\and
\IEEEauthorblockN{2\textsuperscript{nd} Tatsuya Komatsu}
\IEEEauthorblockA{\textit{LY Corporation} \\
Tokyo, Japan \\
komatsu.tatsuya@lycorp.co.jp}
\and
\IEEEauthorblockN{3\textsuperscript{rd} Yoshimitsu Aoki}
\IEEEauthorblockA{\textit{Keio University} \\
Kanagawa, Japan \\
aoki@elec.keio.ac.jp
}
}

\maketitle

\begin{abstract}
In this paper, we propose to pre-train audio encoders using synthetic patterns instead of real audio data.
Our proposed framework consists of two key elements.
The first one is Masked Autoencoder (MAE), a self-supervised learning framework
that learns from reconstructing data from randomly masked counterparts.
MAEs tend to focus on low-level information such as visual patterns and regularities within data.
Therefore, it is unimportant what is portrayed in the input,
whether it be images, audio mel-spectrograms, or even synthetic patterns.
This leads to the second key element, which is synthetic data.
Synthetic data, unlike real audio, is free from privacy and licensing infringement issues.
By combining MAEs and synthetic patterns,
our framework enables the model to learn generalized feature representations without real data,
while addressing the issues related to real audio.
To evaluate the efficacy of our framework,
we conduct extensive experiments across a total of 13 audio tasks and 17 synthetic datasets.
The experiments provide insights into which types of synthetic patterns are effective for audio.
Our results demonstrate that our framework achieves performance
comparable to models pre-trained on AudioSet-2M and partially outperforms image-based pre-training methods.
\end{abstract}

\begin{IEEEkeywords}
self-supervised learning,
masked autoencoder, audio, synthetic data
\end{IEEEkeywords}

\section{Introduction}
\label{sec:introduction}

Large-scale models have demonstrated high performance in the audio processing field.
Among these, Transformers~\cite{vaswani2017attention} have played an important role
in advancing this field,
although they have the drawback of requiring large amounts of labeled data for training.
Moreover, it is challenging to collect high-quality labeled audio data in the real world,
which impedes effective training.
We identify two approaches in existing works aiming to leverage Transformers to solve downstream tasks:
(i) transferring Vision Transformers~\cite{dosovitskiy2020image} (ViTs) pre-trained on ImageNet~\cite{deng2009imagenet}
to audio tasks~\cite{gong2021ast},
(ii) learning feature representations through self-supervised learning~\cite{huang2022masked,
chong2023masked,georgescu2023audiovisual}
from large amounts of audio data (e.g., AudioSet~\cite{gemmeke2017audio})
and VGGSound~\cite{chen2020vggsound}).

However, these methods harbor problems related to real data, as described below:

\noindent \underline{\textbf{Privacy issues}}:
ImageNet and AudioSet consist of images that portray people and human voices, respectively.
Therefore, using these datasets may potentially infringe on privacy.

\noindent \underline{\textbf{License infringement issues}}:
Most of the data included in large-scale datasets are collected from the web,
such as YouTube and search engines.
Some of these data may have licenses that prohibit their use for model training.
Using such data or models trained on such data is legally sensitive.

\begin{figure}[t]
    \centering
    \small
    \includegraphics[width=1.0\linewidth]{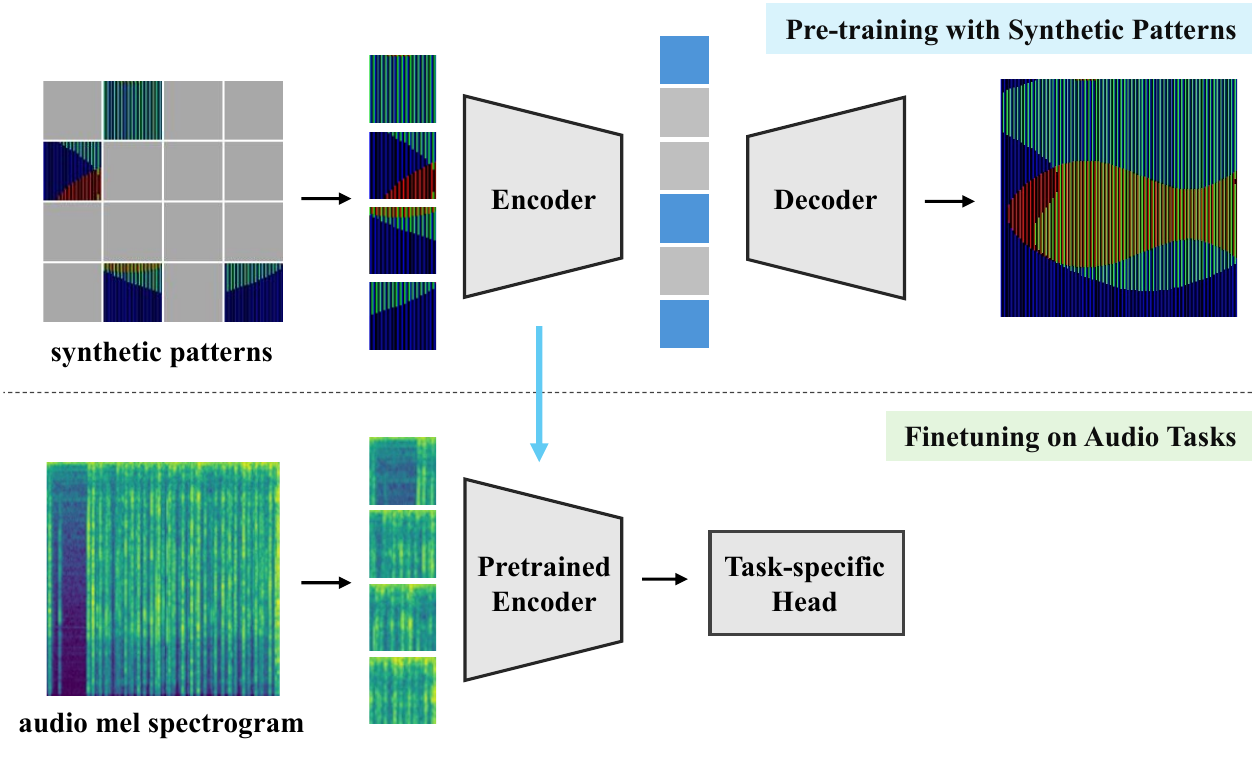}

    \vspace{-1mm}
    \caption{
    \textbf{Overview of our proposed framework.}
    In our framework, we first pre-train a Masked Autoencoder (MAE) using synthetic patterns,
    and then finetune its encoder part for downstream audio tasks.
    This approach eliminates the need for real data during pre-training.
    }
    \vspace{-2mm}
    \label{fig:proposed_method}
\end{figure}

One approach to solving these issues is 
to synthesize realistic data.
In the audio domain,
while some works have proposed utilizing text-to-speech systems~\cite{rosenberg2019speech,
zheng2021using,bartelds2023making},
others have suggested training models
with sounds generated by synthesizers~\cite{engel2017neural,cherep2024contrastive}.
However, many of these approaches still rely on real data
and have subpar performance when using synthetic data alone
primarily due to a lack of diversity.

As an alternative approach in computer vision,
existing works have proposed pre-training models with synthetic visual patterns~\cite{kataoka2020pre,
takashima2023visual,baradad2021learning,baradad2022procedural}.
These methods have been reported to demonstrate high performance in image recognition.
Moreover,~\cite{ishikawa2024data} has shown that VideoMAE~\cite{tong2022videomae},
a Masked Autoencoder (MAE)~\cite{he2022masked} for videos,
can learn spatiotemporal features from videos generated from synthetic images,
although these videos do not contain any humans or realistic objects.
This result implies that MAEs learn domain-agnostic, low-level features like patterns and structures
rather than high-level semantic features such as portrayed objects or actions.

In this paper, inspired by these methods,
we propose to pre-train audio encoders using synthetic patterns,
addressing issues related to privacy and licensing during audio pre-training.
Our framework (Fig.~\ref{fig:proposed_method}) combines two key elements.
The first is an MAE, which is trained to reconstruct the whole input from randomly masked counterparts.
Since MAEs tend to focus on low-level information like visual patterns and regularities within an input,
it is not important to what is portrayed in the input,
whether it be real images, real audio mel-spectrograms, even or synthetic patterns.
This leads to the second key element, which is synthetic data.
Synthetic data, unlike real images and real audio, is free from concerns about privacy and licensing.
By combining MAEs and synthetic patterns,
our framework enables the model to learn transferable feature representations without real audio,
performing well on various audio downstream tasks.

To demonstrate the efficacy of our framework,
we conduct extensive experiments on a total of 13 audio tasks.
In the experiments, we utilize 17 existing synthetic images as synthetic patterns
and evaluate which types of synthetic patterns are effective for audio.
Our experimental results demonstrate that our framework
achieves performance comparable to those pre-trained with AudioSet-2M
and partially surpasses other pre-training methods using images.
These findings suggest that our framework can be a solution to issues related to real data such as privacy and licensing during pre-training,
without impeding performance on audio tasks.

\section{Proposed Framework}
\label{sec:proposed_method}

To address privacy and licensing concerns in pre-training audio encoders,
we propose pre-training an MAE with synthetic patterns
and then transferring it to audio tasks (Fig.~\ref{fig:proposed_method}).
In this section, We first provide an overview of MAEs (Sec.~\ref{subsec:masked_autoencoder}),
and then describe the synthetic patterns used in our experiments (Sec.~\ref{subsec:synthetic_patterns}).
Finally, we explain how to transfer MAEs to audio data (Sec.~\ref{subsec:transferring}).

\newcommand{\red}[1]{\textcolor{red}{#1}}
\newcommand{\blue}[1]{\textcolor{blue}{#1}}
\newcommand{\magenta}[1]{\textcolor{magenta}{#1}}
\newcommand{\cyan}[1]{\textcolor{cyan}{#1}}
\newcommand{\orange}[1]{\textcolor{orange}{#1}}
\newcommand{\yellow}[1]{\textcolor{yellow}{#1}}

\begin{figure*}[t]
    \centering
    \small
    \includegraphics[width=1.0\linewidth]{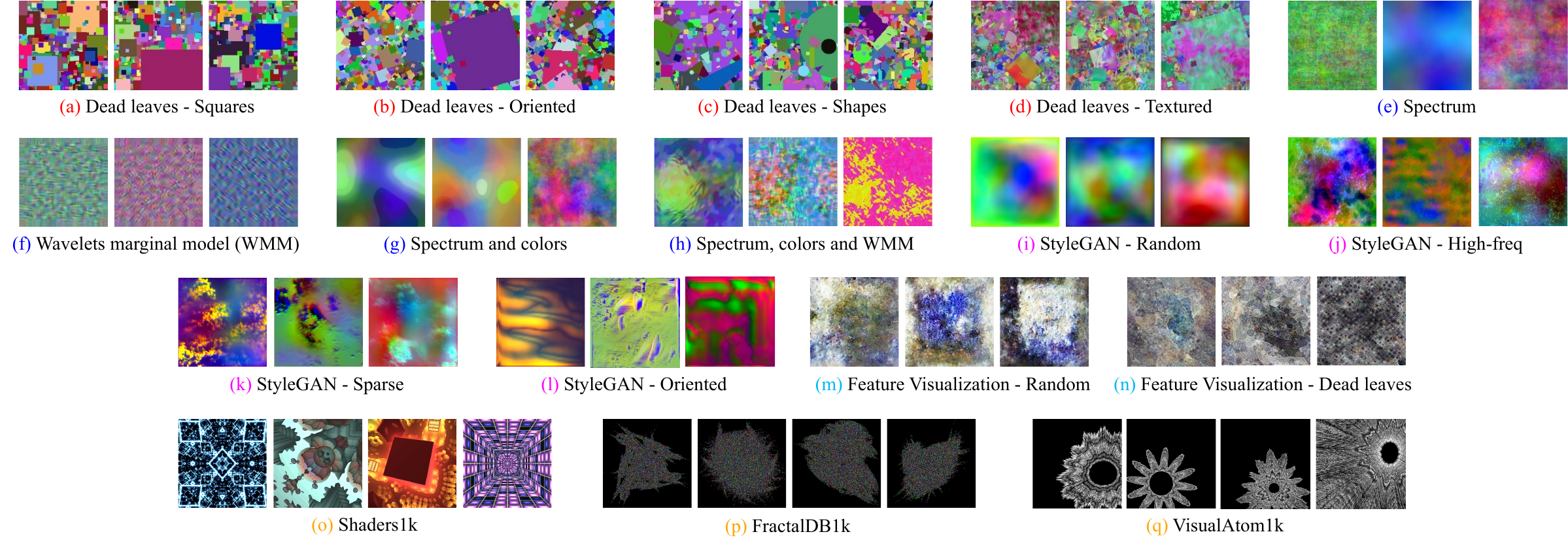}

    \vspace{-1mm}
    \caption{
        \textbf{Examples of synthetic image dataset used in our work.}
        Datasets (a-n) are proposed in ~\cite{baradad2021learning}.
        \red{(a-d)} Dead-leave models,
        \blue{(e-h)} Statistical image models,
        \magenta{(i-l)} StyleGAN-based models,
        and \cyan{(m-n)} Feature visualization.
        Datasets \orange{(o-q)} are large-scale synthetic datasets.
        (o) Shaders1k~\cite{baradad2022procedural}, (p) FractalDB1k~\cite{kataoka2020pre},
        and (q) VisualAtom1k~\cite{takashima2023visual}.
    }
    \vspace{1mm}

    \label{fig:synthetic_images}
\end{figure*}
\begin{figure*}[t]
\centering
\footnotesize

\vspace{-1mm}
\begin{subfigure}[b]{0.195\textwidth}
\centering
\includegraphics[width=\textwidth]{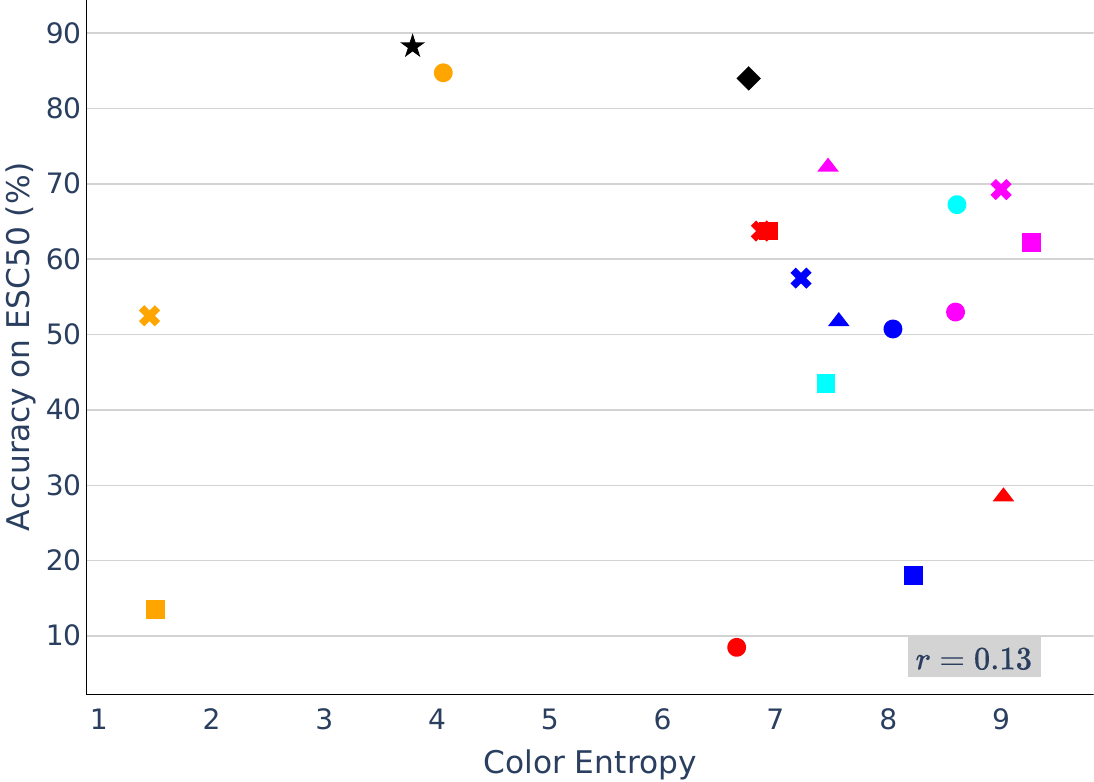}
\caption{Color entropy}
\label{fig:color_vs_acc}
\end{subfigure}
\hfill
\begin{subfigure}[b]{0.195\textwidth}
\centering
\includegraphics[width=\textwidth]{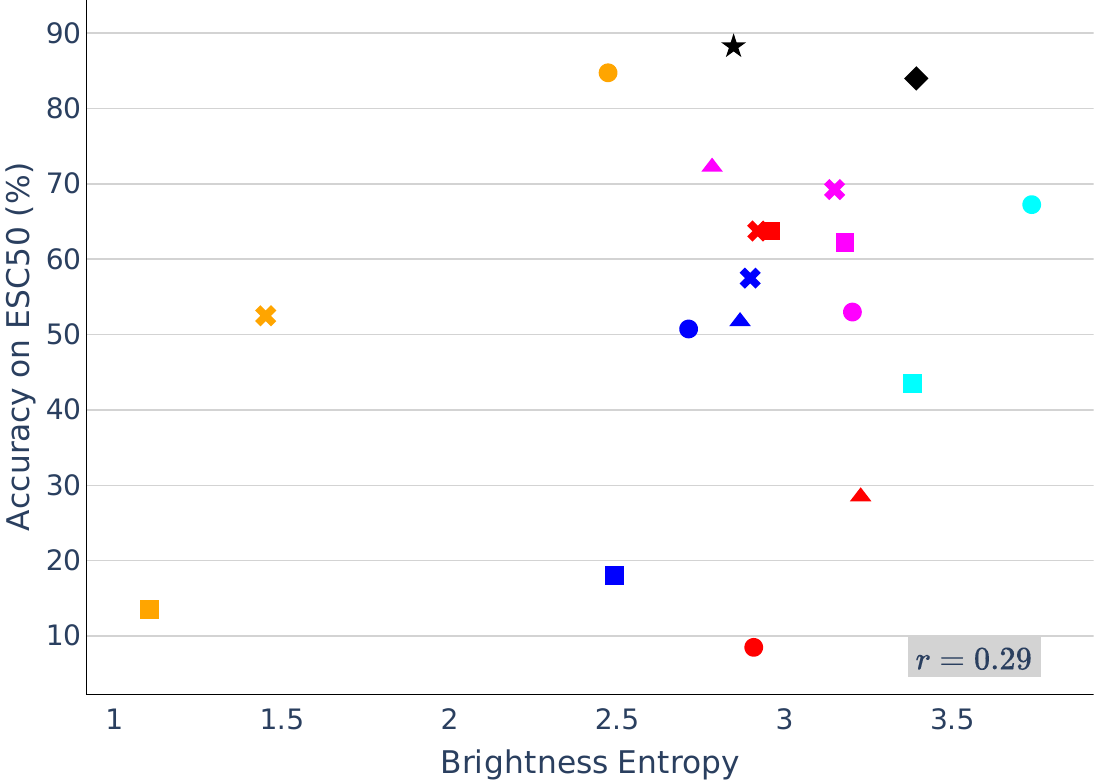}
\caption{Brightness entropy}
\label{fig:brightness_vs_acc}
\end{subfigure}
\hfill
\begin{subfigure}[b]{0.195\textwidth}
\centering
\includegraphics[width=\textwidth]{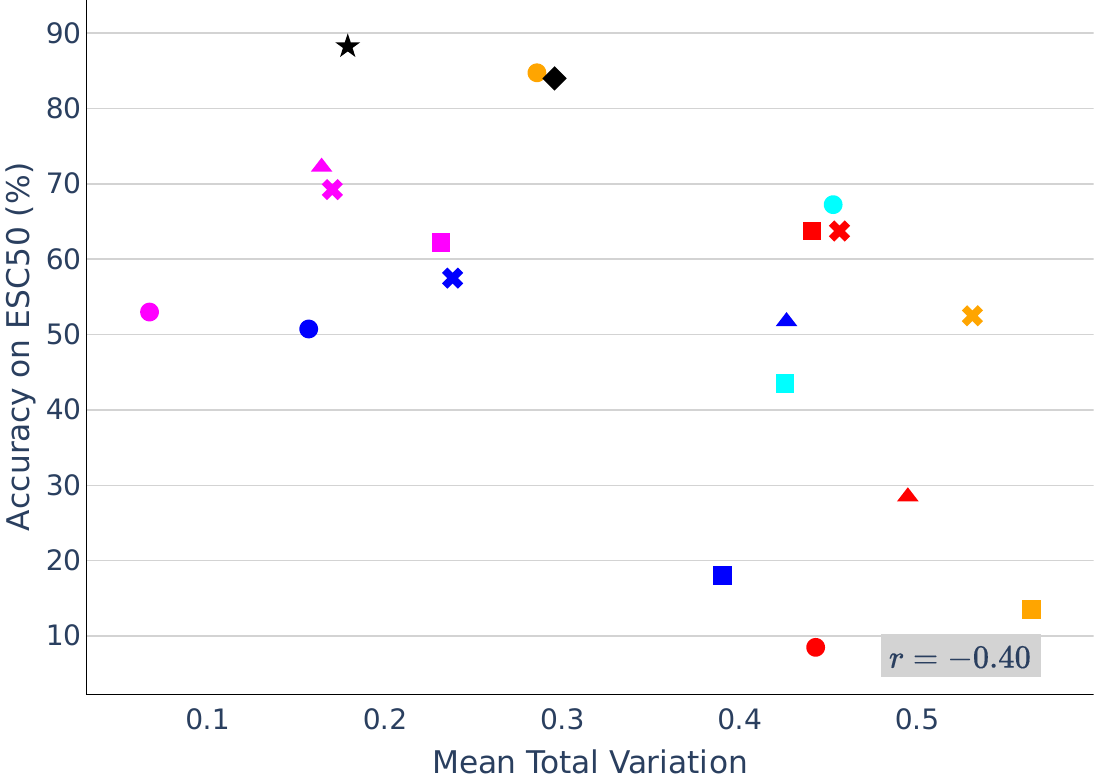}
\caption{Mean total variation}
\label{fig:total_variation_vs_acc}
\end{subfigure}
\hfill
\begin{subfigure}[b]{0.195\textwidth}
\centering
\includegraphics[width=\textwidth]{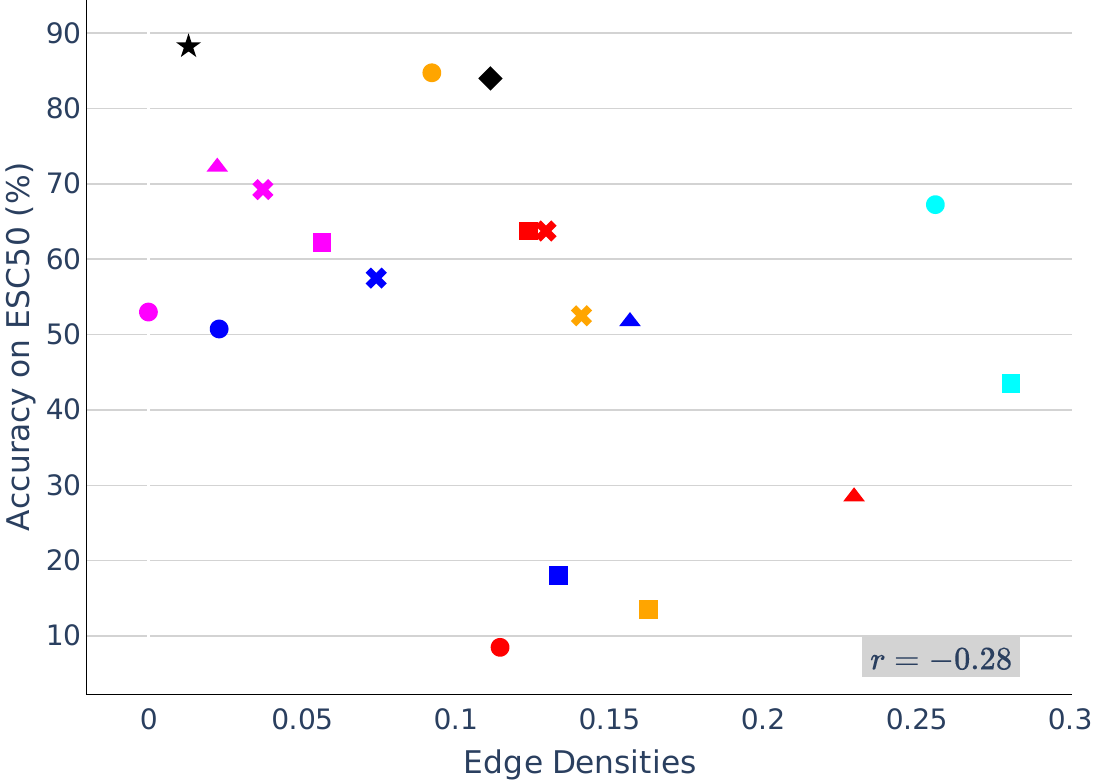}
\caption{Mean edge density}
\label{fig:edge_vs_acc}
\end{subfigure}
\hfill
\begin{subfigure}[b]{0.195\textwidth}
\centering
\includegraphics[width=\textwidth]{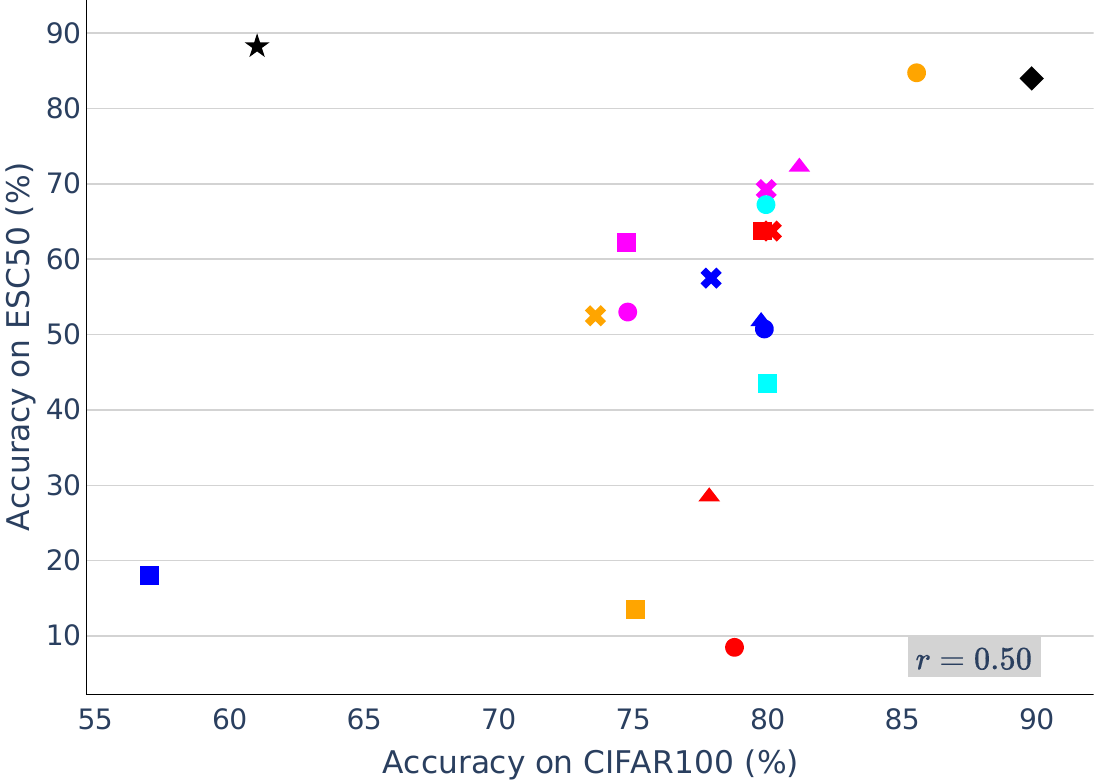}
\caption{Accuracy on CIFAR100}
\label{fig:cifar100_vs_acc}
\end{subfigure}
\\
\vspace{1mm}
\begin{subfigure}[b]{1.0\textwidth}
\centering
\includegraphics[width=\textwidth]{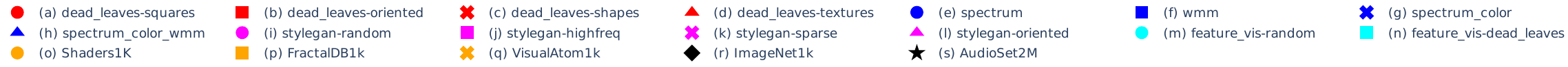}
\label{fig:legend}
\end{subfigure}
\vspace{-5mm}
\caption{
\textbf{Correlation between synthetic image properties and performance on ESC-50 fold 5.}
Note that we use only small-scale datasets (a-n) for calculating the correlation coefficient $r$.
}
\label{fig:ablation_graph}
\end{figure*}

\subsection{Masked Autoencoder}
\label{subsec:masked_autoencoder}

Masked Autoencoder (MAE)~\cite{he2022masked} is a self-supervised learning framework,
where the transformer-based autoencoder aims to reconstruct an input from its masked version.
While MAE was originally developed for image recognition,
it has been successfully applied to audio~\cite{huang2022masked,chong2023masked}.

The MAE pre-training process (illustrated in the upper half of Fig.~\ref{fig:proposed_method})
begins by splitting the input the input $X \in \mathbb{R}^{C \times H \times W}$
\footnote{$C$ represents the number of channels
and $H$ and $W$ each represent the height and width of the input.
}
into non-overlapping $P \times P$ patches,
which are then linearly embedded to form $X_p \in \mathbb{R}^{N \times D}$,
where $N = \frac{HW}{P^2}$ is the number of patches and $D$ is the embedding dimension.
Subsequently, a high proportion of patches (e.g., 75\%) are randomly masked out
using a binary mask, producing visible embeddings.
These embeddings are then fed into a transformer-based autoencoder,
which is trained to reconstruct the original pixel values using mean squared error as the loss function.
After pre-training, the encoder part of the MAE, which is equivalent to a Vision Transformer (ViT),
will be fine-tuned for downstream tasks (in the lower part of Fig.~\ref{fig:proposed_method}).

While existing MAE approaches typically use real-world data,
~\cite{ishikawa2024data} has demonstrated that 
VideoMAE~\cite{tong2022videomae} can learn spatiotemporal feature representations
for action recognition, from synthetic videos which do not contain humans or objects.
This suggests that MAEs focus on low-level information (e.g. visual patterns and regularities)
and it is unimportant what is portrayed in the input, whether it be images, audio mel-spectrograms, or even synthetic patterns.

Inspired by this finding, we propose pre-training MAEs with synthetic patterns,
and then transfering them to audio tasks.
This approach can eliminate the need for real audio data during pre-training,
therefore alleviating issues related to real data like privacy and license infringement.
In the following section, we will elaborate on the synthetic patterns that we use in this study.

\subsection{Synthetic Patterns}
\label{subsec:synthetic_patterns}

While various types of synthetic patterns can be used for training MAEs,
it is challenging to explore effective synthetic patterns for MAEs without any constraints.
Therefore, in this study, we focus on synthetic images as one form of synthetic patterns.
Specifically, we utilize 17 synthetic image datasets proposed in computer vision,
as illustrated in Fig.~\ref{fig:synthetic_images}.

Datasets labeled (a-n) contain 100k images, respectively,
and are generated from structured noise~\cite{baradad2021learning}.
We use these datasets to investigate which types of synthetic patterns
are beneficial for MAE training.
We also use large-scale synthetic image datasets labeled (o-q)
for comparison with existing pre-training methods.
FractalDB1k~\cite{kataoka2020pre} and
VisualAtom1k~\cite{takashima2023visual}
are generated from mathematical formulas,
while Shaders1k~\cite{baradad2022procedural} is synthesized with OpenGL fragment shaders.
Each of these large-scale datasets includes over 1M images.
Pre-training on these datasets achieves performance on image classification tasks
comparable to that achieved by pre-training on large-scale real image datasets such as ImageNet.

\subsection{Transferring MAEs to Audio Tasks}
\label{subsec:transferring}

When we pre-train MAEs with synthetic images and transfer them to audio tasks,
we need to modify its encoder part (i.e., Vision Transformer; ViT)
because the input shape is different between ViTs and Audio Spectrogram Transformers (ASTs)~\cite{gong2021ast}.
We make the following modifications:

\noindent \textbf{Patch Embedding:}
The input dimension $C$ differs between ViTs (3 channels) and ASTs (1 channel).
To enable RGB-image-based ViTs to handle 1-channel inputs,
we sum the weights of the patch embedding along the channel dimension.

\noindent \textbf{Positional encoding:}
The input size of ViTs is typically different from that of ASTs,
while the patch size is the same.
To address this, we directly modify the positional encoding.
Both ViTs and ASTs use sinusoidal positional encoding,
so we simply replace the positional encoding of ViTs with that of ASTs.

\section{Experiments}
\label{sec:experiments}

To evaluate the efficacy of our framework,
we pre-train MAEs using synthetic images
and fine-tune them on various audio tasks.
We provide the details of the experimental setting as follows.

\noindent \textbf{Datasets: }
For pre-training, we use 17 synthetic image datasets
described in Sec.~\ref{subsec:synthetic_patterns}.
For downstream tasks, we use a total of 12 datasets.
Following~\cite{chong2023masked},
we evaluate our framework on
AudioSet-20k/2M (AS-20k/2M)~\cite{gemmeke2017audio},
ESC50~\cite{piczak2015esc},
DCASE2019 task1A dataset~\cite{mesaros2019acoustic},
OpenMIC2018 dataset~\cite{humphrey2018openmic},
and Speech Command V2 (SCV2)~\cite{warden2018speech}.
Additionally, following~\cite{cherep2024contrastive},
we conduct experiments on selected tasks
from HEAR benchmark~\cite{turian2022hear} and ARCH benchmark~\cite{la2024benchmarking}:
UrbanSound 8K (US8K)~\cite{salamon2014dataset},
Variably Intense Vocalizations of Affect and Emotion dataset (VIVAE)~\cite{holz2022variably},
NSynth Pitch 5h dataset (NSynth)~\cite{engel2017neural},
CREMA-D (C-D)~\cite{cao2014crema},
FSD50k~\cite{fonseca2021fsd50k},
Vocal Imitations dataset (VI)~\cite{kim2018vocal},
and LibriCount dataset (LCount)~\cite{fabian_robert_stoter_2018_1216072}.
We report mean average precision (mAP) for AS-20k/2M, OpenMIC2018, FSD50k, and VI,
while we use accuracy as an evaluation metric for other datasets.
For datasets with multi-fold splits,
we conduct cross-validation and report the average metric.

\noindent \textbf{Implementation Details: }
We conducted pre-training MAEs following the setting of ~\cite{he2022masked},
with a mask ratio of 0.75 and the number of epochs set to 800.
For fine-tuning, our experiments are mainly based on MaskSpec~\cite{chong2023masked}
due to high reproducibility.
For the model architecture,
We adopted a vanilla ViT~\cite{dosovitskiy2020image}
with non-overlapped patches as the backbone, especially the ViT-Base variant.

\subsection{Properties of synthetic images}

First, we used small-scale synthetic image datasets (labeled a-n in Fig.~\ref{fig:synthetic_images})
to investigate which types of synthetic images are effective for our framework.
We examined the Pearson correlation coefficient $r$
between the following four properties of the datasets
and their performance on ESC-50 fold5 (Fig.~\ref{fig:ablation_graph}).

\noindent \textbf{Color Entropy:}
To measure the diversity of colors in images, we calculated color histograms for each image and computed the average entropy of these histograms.
Fig.~\ref{fig:color_vs_acc} shows that color diversity has little correlation with performance.

\noindent \textbf{Brightness Entropy:}
We converted images to grayscale, calculated brightness histograms, and computed their average entropy.
We found there is a weak correlation with performance (Fig.~\ref{fig:brightness_vs_acc}).

\noindent \textbf{Total Variation:}
Total Variation (TV)~\cite{rudin1992nonlinear} is typically used as regularization
in image restoration and denoising.
Here, we adopt TV as a metric to evaluate how much noise is in the images
We calculated the sum of TV for each image and reported the average value.
We observed a negative correlation with performance ($r = -0.4$).
This suggests that image datasets with fewer noises and sharp changes of textures
can be more effective for transferring to audio tasks.

\begin{table}[t!]
\footnotesize

\centering
\caption{
\textbf{Comparison of transferability of image-based pre-training methods.}
SL = Supervised Learning, FDSL = Formula-Driven Supervised Learning,
IN-1k = ImageNet1k, exF21k = exFractal21k~\cite{kataoka2022replacing}
VA = VisualAtom~\cite{takashima2023visual}.
$^\ast$ indicates results from~\cite{gong2021ast}.
Note that we are missing 30k data in the balanced and the eval set in AS-20k.
Therefore there is a performance gap between ours and ImageNet1k SL~\cite{gong2021ast}.
However, the gap narrows when the same data is used.
}

\label{tab:compare_pretraining_dataset}
\begin{tabular}{cccccc}

\toprule[1.2pt]
\multicolumn{3}{c}{Pre-training setting}                                & \multicolumn{3}{c}{Downstream tasks} \\
Method                       & Dataset                                  & Labels       & AS-20k    & ESC50   & IN-1k  \\
\midrule[0.5pt]
from scratch$^\ast$          & -                                        &              & 0.148     & -       & 77.9   \\
SL$^\ast$                    & IN-1k                                    & \checkmark   & 0.347     & 0.887   & 83.4    \\
SL                           & IN-1k                                    & \checkmark   & 0.290     & 0.870   & 83.4    \\
FDSL                         & exF-21k                                  &              & 0.236     & 0.837   & 82.7    \\
FDSL                         & VA-21k                                   &              & 0.172     & 0.681   & 83.7    \\
\midrule[0.5pt]
\multirow{4}{*}{MAE}         & IN-1k                                    &              & 0.262     & 0.858   & 83.6    \\
                             & Shaders1k                                &              & 0.274     & 0.873   & 82.1    \\
                             & FractalDB1k                              &              & 0.067     & 0.136   & 77.1    \\
                             & VA-1k                                    &              & 0.110     & 0.795   & 79.9    \\
\bottomrule[1.2pt]

\end{tabular}
\end{table}
\noindent \textbf{Edge Density:}
Using the Canny edge detector~\cite{canny1986computational},
we calculated the average ratio of edge pixels in the images.
As shown in Fig.~\ref{fig:edge_vs_acc}, we found that edge density showed less correlation than TV.

\noindent \textbf{CIFAR100 Performance:}
We also evaluate each model on the image classification task with CIFAR100~\cite{krizhevsky2009learning}.
Note that when we fine-tune the MAE pre-trained on AudioSet2M,
we inflate the weights of patch embedding along the channel dimension to handle RGB images.
As shown in Fig.~\ref{fig:cifar100_vs_acc}, 
models that perform well on CIFAR100 also tend to perform well on ESC50, and vice versa ($r = 0.5$).
Notably, the model pre-trained on AudioSet2M struggles with the image classification tasks.
This suggests that it is difficult to transfer audio models to image tasks,
while image models tend to learn feature representations that are transferable to audio tasks.

These trends were also observed in large-scale datasets.
Notably, ImageNet and Shaders, which have lower average TV values,
demonstrated higher performance compared to VisualAtom and FractalDB,
which contain noises and complex edges.

Based on these observations,
we can conclude images with lower TV values are more effective for MAE pre-training with synthetic patterns.
In other words, synthetic patterns should have less noise and smoother changes in textures for efficient MAE pre-training.

\begin{table*}[t]
\footnotesize
\centering
\caption{
    \textbf{Comparison with other existing methods.}
    IN-1k = ImageNet1k, AS = AudioSet.
}

\label{tab:compare_maskspec_benchmark}
\begin{tabular}{ccccccccccc}
\toprule[1.2pt]
\multirow{2}{*}{Method}           & \multicolumn{3}{c}{Pre-training Setting}                        & \multicolumn{6}{c}{Downstream Tasks}                                            \\ 
                                  & Dataset                                          & Real data  & Label      & AS-2M       & AS-20k       & ESC50   & DCASE2019 & OpenMIC18 & SCV2  \\
\midrule[0.5pt]
CNN14~\cite{kong2020panns}        & -                                                & -          & -          & 0.431       & -            & 0.833   & 0.691     & -         & -       \\
CNN14~\cite{kong2020panns}        & AS                                               & \checkmark & \checkmark & 0.431       & 0.278        & 0.947   & 0.764     & -         & -       \\
PSLA~\cite{gong2021psla}          & AS                                               & \checkmark & \checkmark & 0.443       & 0.319        & 0.877   & -         & -         & -       \\
AST~\cite{gong2021ast}            & IN-1k + AS                                       & \checkmark & \checkmark & 0.457       & 0.347        & 0.956   & -         & -         & 0.981   \\
PaSST~\cite{koutini2021efficient} & IN-1k + AS                                       & \checkmark & \checkmark & 0.471       & -            & 0.968   & -         & 0.843     & -       \\
SSAST~\cite{gong2022ssast}        & Librispeech~\cite{panayotov2015librispeech} + AS & \checkmark &            & -           & 0.310        & 0.888   & -         & -         & 0.980   \\
MaskSpec~\cite{chong2023masked}   & AS                                               & \checkmark &            & 0.471       & 0.323        & 0.896   & 0.801     & 0.814     & 0.977   \\
\midrule[0.5pt]
Ours                              & Shaders1k                                        &            &            & 0.461       & 0.274        & 0.873   & 0.763     & 0.790     & 0.968   \\
\bottomrule[1.2pt]

\vspace{1mm}
\end{tabular}
\end{table*}

\begin{table*}[]
\footnotesize
\centering
\caption{
\textbf{Evaluation on datasets from HEAR and ARCH benchmark.}
$^\ast$ indicates results from~\cite{cherep2024contrastive}.
SL = Supervised Learning.
}
\vspace{-0.1mm}
\label{tab:hear_benchmark}
\begin{tabular}{cccccccccccc}
\toprule[1.2pt]
\multirow{2}{*}{Method}                          & \multicolumn{3}{c}{Pre-training Setting} & \multicolumn{8}{c}{Downstream Tasks}    \\ 
                                                 & Modality    & Real Data  & Label      & ESC50 & US8K  & VIVAE & NSynth& C-D   & FSD50k& VI    & LCount \\
\midrule[0.5pt]
\multicolumn{12}{c}{\textbf{Linear Probing}}                                                                                                                   \\
\midrule[0.5pt]
MS-CLAP~\cite{elizalde2023clap}                  & audio-text  & \checkmark & \checkmark & 0.931 & 0.839 & -     & -     & 0.283 & 0.591 & -     & 0.572    \\
VGGSound SL~\cite{chen2020vggsound}$^\ast$       & audio       & \checkmark & \checkmark & 0.875 & 0.776 & 0.394 & 0.438 & 0.544 & 0.438 & 0.141 & 0.561    \\
VGGSound SSL$^\ast$                              & audio       & \checkmark &            & 0.530 & 0.638 & 0.381 & 0.142 & 0.500 & 0.240 & 0.343 & 0.698    \\
Audio Doppelgängers~\cite{cherep2024contrastive} & audio       &            &            & 0.589 & 0.667 & 0.395 & 0.444 & 0.484 & 0.241 & 0.915 & 0.586    \\
MaskSpec~\cite{chong2023masked}                  & audio       & \checkmark &            & 0.451 & 0.561 & 0.362 & 0.486 & 0.409 & 0.142 & 0.044 & 0.467    \\
ImageNet SL                                      & image       & \checkmark & \checkmark & 0.357 & 0.473 & 0.388 & 0.090 & 0.443 & 0.146 & 0.049 & 0.423    \\
Shaders1k MAE (Ours)                             & image       &            &            & 0.343 & 0.508 & 0.340 & 0.252 & 0.399 & 0.104 & 0.037 & 0.443    \\
\midrule[0.5pt]
\multicolumn{12}{c}{\textbf{Fine-tuning}}                                                                                                                      \\
\midrule[0.5pt]
MaskSpec~\cite{chong2023masked}                  & audio       & \checkmark &            & 0.896 & 0.769 & 0.421 & 0.810 & 0.557 & 0.573 & 0.107 & 0.669    \\
ImageNet SL                                      & image       & \checkmark & \checkmark & 0.870 & 0.776 & 0.467 & 0.798 & 0.581 & 0.573 & 0.135 & 0.679    \\
Shaders1k MAE (Ours)                             & image       &            &            & 0.873 & 0.783 & 0.422 & 0.850 & 0.575 & 0.563 & 0.129 & 0.684    \\
\bottomrule[1.2pt]
\end{tabular}
\end{table*}
\subsection{Comparison with image-based pre-training}

To demonstrate the transferability of MAEs,
we compare MAEs pre-trained on synthetic images
with other image-based pre-training methods.
For this comparison, we use ViT pre-trained with supervised learning on ImageNet1k and Formula-Driven Supervised Learning (FDSL)
with exFractal21k~\cite{kataoka2022replacing} and VisualAtom21k~\cite{takashima2023visual}.
FDSL is a powerful pre-training framework where the model aims to classify images generated from mathematical formulae
into predefined categories.

Table~\ref{tab:compare_pretraining_dataset} shows the results.
Although supervised learning on ImageNet1k achieves superior performance,
it requires real images and high-quality annotations.
In contrast, our framework with Shaders1k enables effective pre-training solely with synthetic images,
thereby alleviating privacy and licensing issues.
Notably, the MAE pre-trained on Shaders1k demonstrates higher performance than those pre-trained on ImageNet1k.
As observed in our previous experiments (Fig.~\ref{fig:ablation_graph}),
pre-training with FractalDB1k and VisualAtom1k also fails to transfer effectively to audio tagging on AudioSet-20k
and environment sound classification on ESC50,
despite their high performance on ImageNet1k.
This indicates that models pre-trained by MAE have higher transferability than models pre-trained by FDSL.
Based on these results, we will use Shaders1k for pre-training in our framework for subsequent experiments.

\subsection{Comparison with existing methods}

Table~\ref{tab:compare_maskspec_benchmark} shows the comparison results
between the proposed framework and existing pre-training approaches on
AS-20k, AS-2M, ESC50, DCASE2019 task1A dataset, OpenMIC18 dataset, and SCV2.
Despite not using audio data during pre-training,
our framework achieves performance that closely approaches that of MaskSpec.
This indicates that the image-based MAE can learn highly transferable features.
While it does not match the performance of supervised learning methods,
these approaches use real data and require high-quality labels,
which incurs high costs of data collection and potentially raises privacy and licensing issues.

\subsection{Evaluation on HEAR and ARCH benchmark}

Table~\ref{tab:hear_benchmark} presents the results from eight datasets
selected from the HEAR and ARCH benchmarks,
following the setting of~\cite{cherep2024contrastive}.
Although our model performs well when fully fine-tuned,
it exhibits a significant limitation in the linear probing setting,
where the encoder weights remain fixed and only the linear layer is trained.
However, this limitation is not unique to our framework; it is also observed in MaskSpec,
which is a self-supervised learning method using MAE and real audio.
Therefore, we consider this limitation results from the characteristics of MAE
which focuses on low-level features, rather than high-level features.
To learn more audio-specific feature representations and improve performance in the linear probing setting,
a more sophisticated approach beyond masked visual modeling is necessary,
which remains a challenge for future work.

\vspace{1mm}
\section{Conclusion}
\label{sec:conclusion}

In this work, we propose pre-training audio encoders utilizing synthetic patterns,
addressing challenges associated with real data, such as privacy concerns and licensing issues.
Our framework demonstrates robust performance across diverse audio tasks, even without audio data during pre-training.
Through extensive experiments,
we have revealed what types of synthetic patterns are effective for audio tasks.
Specifically, we have found that smoother images with fewer Total Variations contribute
significantly to MAE pre-training.
In comparison with existing methods,
our framework achieves comparable performance to self-supervised pre-training methods with real audio
and partially outperforms image-based pre-training methods.
We posit that our framework offers a viable solution
to mitigate the costs of audio data collection and
alleviate concerns regarding privacy and license infringements during audio pre-training.

\bibliographystyle{IEEEtran}
\bibliography{IEEEabrv,refs}

\end{document}